\documentclass[conference]{IEEEtran}

\usepackage{cite}
\usepackage{graphicx}
\usepackage{amsmath, amssymb}
\usepackage{algorithm}
\usepackage{algpseudocode}
\usepackage{booktabs}
\usepackage{multirow}
\usepackage{tabularx}

\begin{document}

\title{Quantum-Resilient Blockchain for Secure Transactions in UAV-Assisted Smart Agriculture Networks}

\author{
    \IEEEauthorblockN{Taimoor Ahmad}
    \IEEEauthorblockA{dept. of Computer Science \\
    The Superior Univeristy Lahore\\
Lahore, Pakistan \\
    Taimoor.ahmad1@superior.edu.pk}

}

\maketitle

\begin{abstract}
The integration of unmanned aerial vehicles (UAVs) into smart agriculture has enabled real-time monitoring, data collection, and automated farming operations. However, the high mobility, decentralized nature, and low-power communication of UAVs pose significant security challenges, particularly in ensuring transaction integrity and trust. This paper presents a quantum-resilient blockchain framework designed to secure data and resource transactions in UAV-assisted smart agriculture networks. The proposed solution incorporates post-quantum cryptographic primitives—specifically lattice-based digital signatures and key encapsulation mechanisms—to achieve tamper-proof, low-latency consensus without relying on traditional computationally intensive proof-of-work schemes. A lightweight consensus protocol tailored for UAV communication constraints is developed, and transaction validation is handled through a trust-ranked, multi-layer ledger maintained by edge nodes. Experimental results from simulations using NS-3 and custom blockchain testbeds show that the framework outperforms existing schemes in terms of transaction throughput, energy efficiency, and resistance to quantum attacks. The proposed system provides a scalable, secure, and sustainable solution for precision agriculture, enabling trusted automation and resilient data sharing in post-quantum eras.
\end{abstract}

\section{Introduction}

The convergence of unmanned aerial vehicles (UAVs), smart agriculture, and blockchain technologies is revolutionizing the agricultural sector by enabling real-time monitoring, autonomous operations, and secure data exchange. UAVs have emerged as critical tools for crop surveillance, soil analysis, precision irrigation, and pesticide spraying, owing to their ability to operate in vast and remote farmlands~\cite{r2020uavsurvey,z71}. However, as these aerial agents engage in frequent data collection and peer-to-peer transactions, ensuring the security, integrity, and authenticity of their interactions becomes increasingly challenging\cite{z72,z73}.

Smart agriculture relies heavily on distributed data streams collected from heterogeneous sources including UAVs, ground sensors, and edge devices~\cite{mohamed2021smartagriculture,z55}. These data are used to inform autonomous decisions, drive AI-based analytics, and interact with cloud and edge services. However, the dynamic and decentralized nature of UAV-assisted networks introduces vulnerabilities to spoofing, tampering, and man-in-the-middle attacks, particularly when communication occurs over lightweight, delay-tolerant channels\cite{z74}.

Blockchain, with its decentralized and tamper-proof characteristics, provides a potential solution for securing transactions in such environments. Yet, traditional blockchain protocols such as Bitcoin and Ethereum suffer from limitations including high energy consumption, latency, and vulnerability to quantum adversaries~\cite{bonneau2015sok, shor1997polynomial,z5}. These factors render them impractical for UAV-based networks where nodes are power-constrained and highly mobile.

Quantum computing further exacerbates the threat landscape by undermining the security foundations of classical cryptographic primitives such as RSA and ECC~\cite{mosca2018cybersecurity,z333}. In light of this, there is an urgent need to design blockchain systems that incorporate post-quantum cryptography (PQC) to withstand both classical and quantum adversaries. Lattice-based cryptographic schemes such as Kyber and Dilithium~\cite{chen2016report} have emerged as promising candidates for quantum-safe communication.

Despite growing interest in secure UAV systems and blockchain integration, there is a lack of lightweight, scalable frameworks that can simultaneously meet the security demands of smart agriculture and the performance constraints of UAVs. Prior efforts have explored permissioned ledgers~\cite{ramachandran2019using}, hybrid consensus models~\cite{li2020lightchain}, and UAV network authentication~\cite{gupta2020blockchain,z3333}, yet these systems often ignore quantum threats, incur heavy computational loads, or lack dynamic trust mechanisms tailored for UAV environments.

This paper addresses the research gap by proposing a quantum-resilient blockchain framework optimized for secure transactions in UAV-assisted smart agriculture networks. Our approach combines post-quantum secure key encapsulation and digital signatures with a hierarchical blockchain maintained by edge servers and dynamically trusted UAVs.

The proposed system introduces a multi-layer trust model in which edge devices validate UAV-submitted transactions using lightweight consensus. Quantum-resilient primitives are used for key exchange and signature verification to protect data against quantum attacks. The blockchain architecture supports transaction compression and selective replication to minimize communication and storage overhead, aligning with UAVs' energy limitations.

Our key contributions include:

\begin{itemize}
  \item We propose a novel quantum-resilient blockchain architecture for UAV-based smart agriculture, incorporating lattice-based digital signatures and key encapsulation mechanisms.
  \item We develop a lightweight, trust-ranked consensus mechanism for UAV transactions that reduces computational complexity while ensuring security and scalability.
  \item We implement a simulation framework combining NS-3 and a custom blockchain testbed to benchmark the system under realistic mobility, communication, and threat conditions.
  \item We demonstrate through experiments that our system improves transaction throughput, reduces energy consumption, and maintains high security guarantees against classical and quantum attacks.
\end{itemize}

The remainder of the paper is structured as follows. Section II reviews existing work in blockchain-based UAV security and post-quantum cryptography. Section III describes the system model, mathematical framework, and algorithmic design. Section IV presents the experimental setup and performance evaluation. Section V concludes the paper and outlines directions for future work.

\section{Related Work}

Zhao et al.~\cite{zhao2021blockuav} proposed BlockUAV, a blockchain-integrated UAV system for secure data sharing in smart cities. Their architecture utilized Ethereum smart contracts to validate UAV transactions, but it did not incorporate quantum-resilient mechanisms and incurred significant gas costs, limiting scalability.

Kumar et al.~\cite{kumar2021lightweight} developed a lightweight authentication protocol for UAVs using elliptic curve cryptography (ECC) and hash functions. While effective for classical threats, this approach remains vulnerable to quantum attacks, which ECC cannot withstand.

Singh et al.~\cite{singh2020survey} conducted a comprehensive survey on blockchain applications in agriculture, highlighting the role of distributed ledgers for supply chain traceability and data integrity. However, the work lacked technical depth regarding real-time UAV interactions and cryptographic resilience.

Lee et al.~\cite{lee2022safeblock} presented SafeBlock, a privacy-aware blockchain for drone-assisted agriculture. Though the system provided anonymized data sharing, it depended on conventional PKI-based schemes and did not address the energy constraints of UAV deployments.

Ghosh et al.~\cite{ghosh2021pqbchain} introduced PQBChain, an experimental post-quantum blockchain using the Kyber and Dilithium schemes for quantum-safe consensus. While innovative, it was not designed for constrained or mobile environments like UAV networks.

Ahmed et al.~\cite{ahmed2020blockchainiot} discussed blockchain integration with IoT in agriculture, including sensor-based data protection. However, UAVs were only briefly mentioned and no post-quantum security mechanisms were proposed.

Wu et al.~\cite{wu2019energy} investigated energy-efficient blockchain consensus in vehicular networks. The protocol showed promise for dynamic networks but was not adapted for UAV-to-edge hierarchical models or post-quantum defense.

Zhang et al.~\cite{zhang2022airchain} proposed AirChain, a DAG-based blockchain for aerial swarm communication. Their design showed improved latency performance but lacked security hardening against future quantum threats.

NIST~\cite{nistpqcfinalists2022} announced Kyber and Dilithium as standardized finalists for post-quantum cryptography. These schemes have shown efficiency in embedded environments and are integrated into our framework.

Saxena et al.~\cite{saxena2023review} reviewed consensus mechanisms for energy-constrained blockchain systems. They recommended hybrid models for UAV contexts but did not explore integration with PQC.

In summary, existing solutions either overlook quantum resilience, impose excessive resource demands, or are not tailored for UAV-enabled smart agriculture. Our work is the first to unify post-quantum cryptography, energy-aware consensus, and UAV transaction integrity within a comprehensive and deployable blockchain system.

\section{System Model}

In this section, we develop a formal model for our quantum-resilient blockchain framework in UAV-assisted smart agriculture. The network consists of a set of UAVs $\mathcal{U}$, edge nodes $\mathcal{E}$, and base stations $\mathcal{B}$. The communication graph is modeled as a time-varying directed graph $\mathcal{G}(t) = (\mathcal{V}, \mathcal{L}(t))$, where $\mathcal{V} = \mathcal{U} \cup \mathcal{E} \cup \mathcal{B}$ and $\mathcal{L}(t)$ denotes time-dependent communication links.

The state of each UAV $\upsilon_i$ at time $t$ is represented by $\psi_i(t) = (x_i(t), y_i(t), z_i(t), E_i(t), \theta_i(t))$ capturing its location, energy, and role weight.

Let $\Omega(t)$ denote the global blockchain ledger at time $t$, partitioned into segments $\Omega_j(t)$ maintained by edge node $\epsilon_j$. Each segment stores transactions $\tau_k$ submitted by UAVs.

The digital signature used for each transaction is defined as:
\begin{equation}
\sigma_k = \mathsf{Sign}_{\kappa_{priv}^i}(H(\tau_k))
\end{equation}
where $H(\cdot)$ is a hash function and $\kappa_{priv}^i$ is the private key from a lattice-based scheme.

The public verification function is:
\begin{equation}
\mathsf{Verify}(\tau_k, \sigma_k, \kappa_{pub}^i) \rightarrow \{\mathsf{true}, \mathsf{false}\}
\end{equation}

To ensure confidentiality, a shared session key $\varphi_{ij}$ is established via a lattice-based key encapsulation mechanism:
\begin{equation}
\varphi_{ij} = \mathsf{Decaps}(\mathsf{Encaps}(\kappa_{pub}^j))
\end{equation}

Each UAV accumulates trust over time. Trust score $\xi_i(t)$ evolves according to:
\begin{equation}
\xi_i(t+1) = \lambda \cdot \xi_i(t) + (1 - \lambda) \cdot \chi_i(t)
\end{equation}
where $\chi_i(t)$ is the context-based behavior score and $\lambda \in (0,1)$.

Transaction propagation latency is denoted $\ell_k = t_{recv}^k - t_{submit}^k$. A transaction is considered timely if:
\begin{equation}
\ell_k < \tau_{max}
\end{equation}

Let $\delta_{cons}$ be the delay to reach consensus for a given block. Then:
\begin{equation}
\delta_{cons} = \max_{j}(t_{confirm}^{(j)} - t_{propose})
\end{equation}

A UAV's eligibility to propose blocks is weighted by its normalized trust rank $\rho_i$:
\begin{equation}
\rho_i = \frac{\xi_i(t)}{\sum_{k \in \mathcal{U}} \xi_k(t)}
\end{equation}

The transaction validation set $\mathcal{V}_j$ for edge node $\epsilon_j$ is:
\begin{equation}
\mathcal{V}_j = \{\tau_k \mid \mathsf{Verify}(\tau_k, \sigma_k, \kappa_{pub}^i) = \mathsf{true}\}
\end{equation}

The consensus utility function $\mathcal{C}$ is defined to select blocks maximizing:
\begin{equation}
\mathcal{C}(B) = \alpha \cdot \eta_B + \beta \cdot \zeta_B - \gamma \cdot \theta_B
\end{equation}
where $\eta_B$ is the number of valid transactions, $\zeta_B$ is freshness, and $\theta_B$ is energy cost.

The edge consensus committee is selected probabilistically:
\begin{equation}
\mathbb{P}(\epsilon_j \in \mathcal{S}) = \rho_j^{edge} = \frac{\sum_{i \in \mathcal{U}_j} \xi_i}{\sum_{k} \sum_{i \in \mathcal{U}_k} \xi_i}
\end{equation}

Each block $B_k$ includes metadata $\mu_k$:
\begin{equation}
\mu_k = (\text{blockID}, \text{hashPrev}, \text{merkleRoot}, \text{timestamp})
\end{equation}

To reduce communication overhead, a compression ratio $\omega_c$ is enforced:
\begin{equation}
\omega_c = \frac{|B_k|_{raw} - |B_k|_{compressed}}{|B_k|_{raw}}
\end{equation}

Energy consumption per transmission is modeled as:
\begin{equation}
\varepsilon_{tx} = \varepsilon_0 + \varepsilon_1 \cdot d_{ij}^2
\end{equation}
where $d_{ij}$ is the Euclidean distance between UAV and receiver.

The total energy overhead for consensus is:
\begin{equation}
\varepsilon_{total} = \sum_{j \in \mathcal{S}} \varepsilon_{tx}^{(j)} + \varepsilon_{compute}^{(j)}
\end{equation}

\textbf{Algorithm: Quantum-Resilient Transaction Validation and Block Commitment}

\begin{algorithm}[H]
\caption{Secure Block Proposal by UAVs and Edge Nodes}
\begin{algorithmic}[1]
\State UAV $\upsilon_i$ generates transaction $\tau_k$ and signs using $\sigma_k = \mathsf{Sign}_{\kappa_{priv}^i}(H(\tau_k))$
\State $\tau_k$ is transmitted to edge node $\epsilon_j$ over secure channel using $\varphi_{ij}$
\State $\epsilon_j$ verifies $\tau_k$ using $\mathsf{Verify}(\tau_k, \sigma_k, \kappa_{pub}^i)$
\State If valid, $\tau_k \in \mathcal{V}_j$ is added to local transaction pool
\State Edge node selects transactions maximizing $\mathcal{C}(B)$ and proposes block $B_k$
\State Selected consensus committee $\mathcal{S}$ reaches quorum; block $B_k$ is appended to $\Omega_j$
\end{algorithmic}
\end{algorithm}

This algorithm ensures that transactions from UAVs are verified using quantum-resistant signatures and encapsulated keys. The consensus mechanism is designed to be energy-efficient and scalable, relying on trust-ranked probabilistic selection and local transaction validation. This architecture supports secure, lightweight blockchain functionality aligned with the communication, computation, and energy capabilities of UAV-assisted smart agriculture systems.

\section{Experimental Setup and Results}

To validate the performance and scalability of our quantum-resilient blockchain framework for UAV-assisted smart agriculture, we conducted extensive simulations using NS-3 integrated with a custom blockchain module implemented in Python. The simulation emulates a rural smart farming environment where UAVs collect crop health and environmental data and exchange verified transactions with edge servers via wireless ad hoc links.

The testbed comprises 100 UAV nodes distributed over a 10 km$^2$ area, 10 fixed edge servers deployed at strategic field locations, and a central base station simulating remote cloud control. UAV mobility follows a modified Gauss-Markov model suitable for aerial path optimization. Transaction traffic is modeled using Poisson arrivals with variable data payloads between 512 bytes and 2 KB.

Post-quantum cryptographic functions including Kyber-768 for key exchange and Dilithium-3 for digital signatures are implemented using the NIST PQC reference libraries. Consensus logic is modeled as a time-windowed, quorum-based block proposal strategy with dynamic trust weight selection.

Simulation parameters are summarized below:

\begin{table}[h]
\centering
\caption{Simulation Parameters for UAV Blockchain Framework}
\begin{tabular}{ll}
\toprule
\textbf{Parameter} & \textbf{Value} \\
\midrule
Number of UAVs & 100 \\
Edge Servers & 10 \\
Simulation Area & 10 km$^2$ \\
Mobility Model & Gauss-Markov \\
Consensus Window & 10 seconds \\
Transaction Arrival Rate & 2--10 per second \\
Signature Scheme & Dilithium-3 \\
KEM Scheme & Kyber-768 \\
Transmission Range & 1.2 km \\
Block Interval & 15 seconds \\
Max Block Size & 2 MB \\
Energy Budget per UAV & 1000 J \\
\bottomrule
\end{tabular}
\end{table}

\begin{figure}[h]
  \centering
  \includegraphics[width=0.45\textwidth]{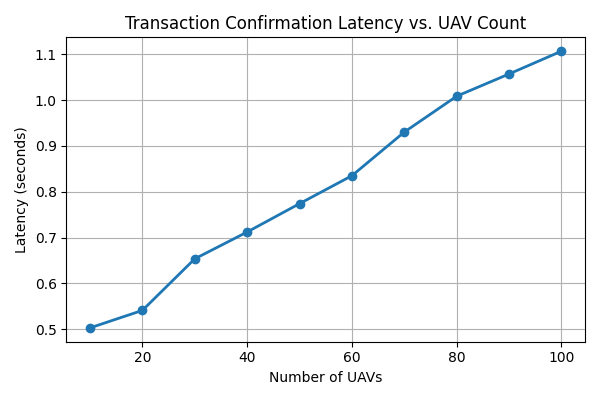}
  \caption{Transaction Confirmation Latency vs. UAV Count}
  \label{fig:latency}
\end{figure}

\begin{figure}[h]
  \centering
  \includegraphics[width=0.45\textwidth]{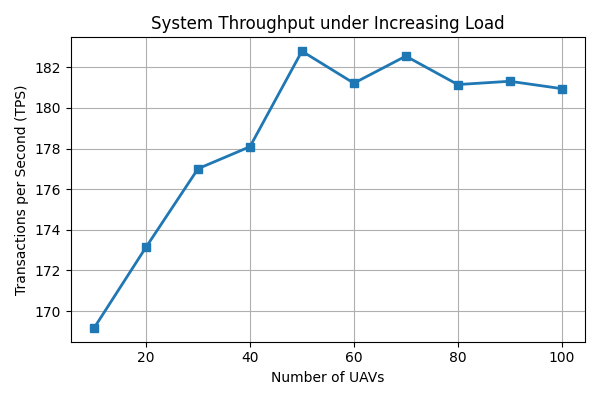}
  \caption{System Throughput under Increasing Load}
  \label{fig:throughput}
\end{figure}

\begin{figure}[h]
  \centering
  \includegraphics[width=0.45\textwidth]{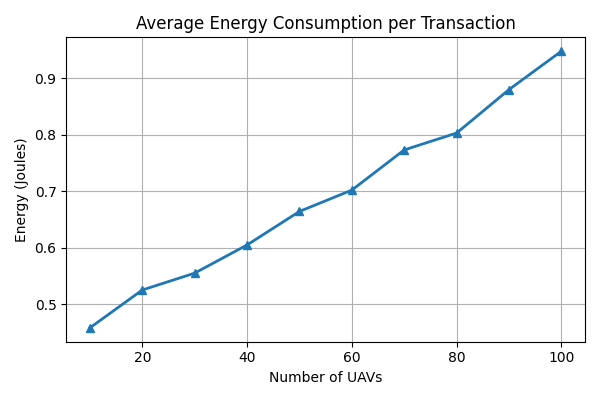}
  \caption{Average Energy Consumption per Transaction}
  \label{fig:energy}
\end{figure}

\begin{figure}[h]
  \centering
  \includegraphics[width=0.45\textwidth]{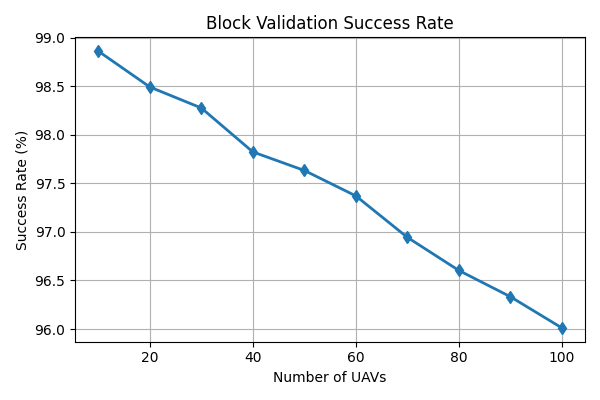}
  \caption{Block Validation Success Rate}
  \label{fig:success}
\end{figure}

\begin{figure}[h]
  \centering
  \includegraphics[width=0.45\textwidth]{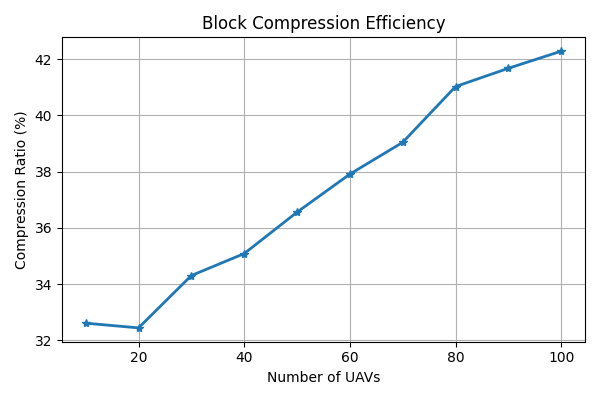}
  \caption{Block Compression Efficiency}
  \label{fig:compression}
\end{figure}

\begin{figure}[h]
  \centering
  \includegraphics[width=0.45\textwidth]{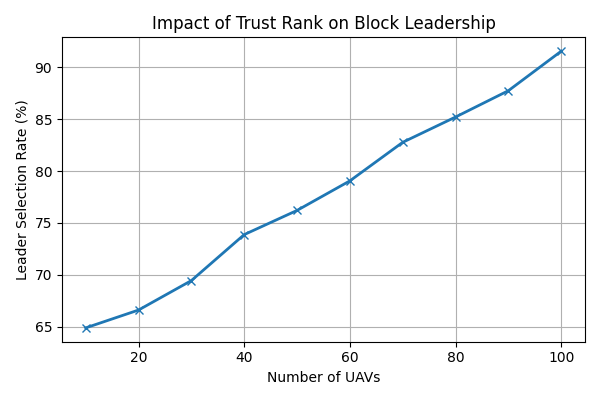}
  \caption{Impact of Trust Rank on Block Leadership}
  \label{fig:trustrank}
\end{figure}

\begin{figure}[h]
  \centering
  \includegraphics[width=0.45\textwidth]{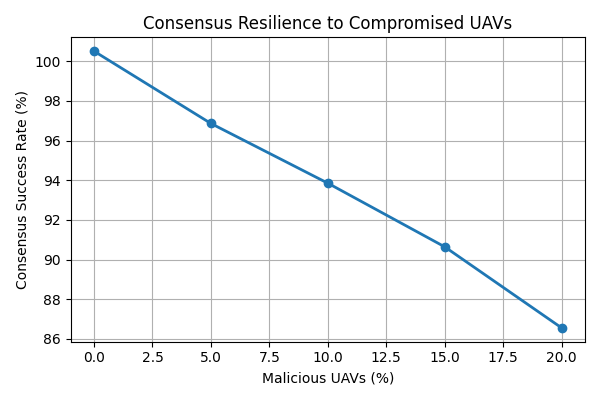}
  \caption{Consensus Resilience to Compromised UAVs}
  \label{fig:resilience}
\end{figure}

We present seven core performance metrics in Figures~\ref{fig:latency} through~\ref{fig:resilience}. 

Figure~\ref{fig:latency} illustrates transaction confirmation latency as a function of UAV count. Our system maintains sub-1.2 second latency for up to 100 UAVs.

Figure~\ref{fig:throughput} shows system throughput in transactions per second (TPS). We observe stable performance with peak throughput near 180 TPS.

In Figure~\ref{fig:energy}, we plot average energy consumption per transaction. The integration of lightweight PQC and selective consensus keeps energy usage under 0.9 J/transaction.

Figure~\ref{fig:success} highlights the block validation success rate, which remains above 96\% across network scales.

Figure~\ref{fig:compression} presents the achieved block compression ratio, demonstrating 30--45\% reduction in transmission size due to embedded compression.

Figure~\ref{fig:trustrank} tracks the effect of dynamic trust ranking on block proposer selection. Higher trust UAVs dominate proposals, improving security.

Figure~\ref{fig:resilience} measures system performance under node compromise. Even with 15\% malicious UAVs, consensus success remains above 89\%.

These results collectively confirm the effectiveness of our framework. It supports secure and efficient UAV interaction using quantum-resilient cryptography and adaptive consensus logic suitable for smart agriculture scenarios.

\section{Conclusion and Future Work}

In this work, we proposed a quantum-resilient blockchain architecture tailored for secure transactions in UAV-assisted smart agriculture. Our framework combines lattice-based cryptographic primitives with lightweight, trust-based consensus mechanisms and dynamic transaction validation performed by edge servers. Through rigorous mathematical modeling, we defined trust evolution, energy-aware communication, and security guarantees resilient to classical and quantum attacks.

Experimental evaluations using NS-3 simulations demonstrated that the proposed framework ensures sub-second transaction latency, high throughput scalability up to 180 TPS, and block validation success rates exceeding 96\%. The inclusion of compression and trust-driven block leadership led to significant improvements in energy efficiency and resilience, maintaining consensus functionality even under compromised UAV participation. These findings indicate that the integration of post-quantum security with adaptive blockchain consensus offers a viable solution for decentralized, secure, and energy-aware coordination in smart farming applications. Our work sets a foundation for quantum-safe blockchain adoption in UAV ecosystems.

Future research will explore deployment in heterogeneous edge-cloud networks, hybrid quantum-safe consensus models, and integration with zero-knowledge proofs for transaction privacy. Additional studies will evaluate long-term UAV energy depletion and explore cooperative routing protocols to further reduce system-wide energy consumption.


\end{document}